\documentclass[doublecol]{epl2} 
\usepackage{amsmath}

\title{Nonlinear diffusion of fermions and bosons}
\shorttitle{Nonlinear diffusion of fermions and bosons} %Insert here a short version of the title if it exceeds 70 characters

\author{Georg Wolschin}
\shortauthor{G. Wolschin}

\institute{                   
Institute for Theoretical Physics, Heidelberg University, Philosophenweg 16, Heidelberg, 69120, Germany
}

\abstract
{A nonlinear diffusion equation is proposed to account for thermalization in fermionic and bosonic systems through analytical solutions.
For constant transport coefficients, exact time-dependent solutions are derived through nonlinear transformations, and the corresponding local equilibration times are deduced.
Fermi--Dirac and Bose--Einstein distributions emerge as stationary solutions of the nonlinear equation. As examples, local thermalization of quarks and gluons in relativistic heavy-ion collisions, and of ultracold atoms including time-dependent Bose--Einstein condensate formation are discussed.}

\begin{document}

\maketitle

\section{Introduction}
\label{intro}
The time-dependent approach of an isolated quantum many-body system from an initial nonequilibrium state to thermal equilibrium
is an outstanding problem in many areas of physics. Examples at opposing ends of the energy scale are the local equilibration of quarks and gluons in the initial stages of relativistic heavy-ion collisions \cite{ba01}, and the thermalization of ultracold atoms, which can be accompanied by the time-dependent formation of a Bose--Einstein condensate (BEC) \cite{miesner_bosonic_1998}. Investigations of thermalization in relativistic heavy-ion collisions at energies reached at the Large Hadron Collider (LHC) have shown that the system has enough time to equilibrate locally. These results emerge from effective kinetic theories that rely on the quantum Boltzmann equation \cite{ba01,km11,jpb12,kur14,jpbl15,blmt17,fuku17}. 

As a complementary mesoscopic approach, a nonlinear diffusion model has been developed \cite{gw18,bgw19,gw20,rgw20,sgw21,gw22} and is now used to calculate closed-form time-dependent fermionic and bosonic occupation-number distributions. The local equilibration of quarks and gluons in high-energy hadronic collisions, as well as the thermalization of bosonic ultracold atoms with time-dependent BEC formation is accounted for. A direct comparison with recent condensate-formation data for $^{39}$K from Cambridge University \cite{gli21} for various scattering lengths underlines the validity and usefulness of the analytical model for both infrared and ultraviolet dynamics. 

The aim is to apply the nonlinear diffusion model \cite{gw18,gw22} in the high-energy regime to the thermalization of quarks and gluons as the relevant partonic constituents in relativistic heavy-ion collisions, and -- in the low-energy region -- to the thermalization of ultracold atoms, including the simultaneous time-dependent condensate formation. Only in the latter case is a direct comparison with experimental data possible, and it will be performed using results \cite{gli21} for ultracold potassium for various scattering lengths.
The kinetic nonlinear diffusion model will thus account for thermalization in both, fermionic and bosonic systems with the proper Fermi--Dirac and Bose--Einstein equilibrium limits. It is solved exactly through a nonlinear transformation, and applied to the local equilibration with suitably adapted transport coefficients in the high- and low-energy regions. The analytical model is complementary to, but simpler than existing numerical solutions of quantum Boltzmann equations. 

Substantial efforts have been made to determine whether a transient gluon condensate can be formed in the initial stages of relativistic collision through elastic number-conserving gluon scatterings \cite{jpb12,jpb13}, bearing similarity to the BEC-formation in ultracold atoms below the critical temperature. It turns out, however, that inelastic number-changing collisions are much faster \cite{blmt17}, hindering condensate formation. In the present work, only inelastic gluon scatterings will be considered.

In contrast, thermalization in ultracold atoms occurs predominantly via number-conserving elastic collisions, triggering Bose-Einstein condensate formation \cite{an95,ket95,hul95,hul97}, and offering the possibility to measure the detailed time-dependence of condensate formation \cite{miesner_bosonic_1998,kdg02,gli21}. Some of these measurements have already been compared with numerical models for the time evolution of the growing condensate \cite{gz97,bzs00,kdg02}. The time-dependent condensate formation of ultracold $^{39}$K atoms that has recently been measured {using} 
%at Cambridge University for various interaction strengths via 
adjustable scattering lengths {\cite{gli21}} will be calculated and interpreted based on the analytical results of the nonlinear diffusion model. 

\section{The model}
\label{model}
The nonlinear diffusion  model is formulated using the ergodic approximation \cite{snowo89,svi91,setk95,lrw96,jgz97}: The collision term for the single-particle 
%expectation-value 
occupation-number distribution depends only on the energy, and on time. The time evolution of the distribution function 
%$\epsilon_\alpha$, 
$n\equiv n_\alpha \equiv \langle n(\epsilon_\alpha,t)\rangle$  is governed by a nonlinear partial differential equation \cite{gw18,gw22} 
 \begin{equation}
%\frac{\partial n}{\partial t}=-\frac{\partial}{\partial\epsilon}\Bigl[v\,n\,(1\pm n)+n\frac{\partial D}{\partial \epsilon}\Bigr]+\frac{\partial^2}{\partial\epsilon^2}\bigl[D\,n\bigr]\,,
\partial_t {n}_{t}=-\partial_\epsilon\bigl[v\,n\,(1\pm n)+n\,\partial_\epsilon D\bigr]+\partial_{\epsilon\epsilon}\bigl[D\,n\bigr]\,,
 \label{boseq}
\end{equation}
%where the $``+"$ sign gives rise to the bosonic, the $``-"$ sign to the fermionic solution. 
where the $+$ sign gives rise to the bosonic, the $-$ sign to the fermionic solution. 
In this nonlinear diffusion equation, the drift term $v\,(\epsilon,t)$ accounts for dissipative effects, whereas $D\,(\epsilon,t)$ mediates diffusion of particles in the energy space.
The many-body physics is contained in these
transport coefficients, which depend on energy, time, and the second moment of the interaction. 

{The equation accounts for the time-dependent 
thermalization in the cloud with particle number $N_\mathrm{th}(t)$, whereas condensate formation via elastic collisions will be considered through particle-number conservation at each timestep, 
$N_\mathrm{c }(t) = N_\mathrm{i} – N_\mathrm{th}(t)$. The effect of interactions between particle cloud and condensate is included indirectly through the choice of the mesoscopic 
transport coefficients $D, v.$

Alternatively, one could consider the condensate via an additional delta-function term in Eq.\,(1), as proposed for quantum Boltzmann equations \cite{jpbl15}. The modified diffusion equation would then conserve the particle number by itself. However, since for cold quantum gases one is interested in the time-dependent particle number in the condensate, the present approach without delta-function term seems more straightforward.}

The stationary solution $n_\infty(\epsilon)$ of the nonlinear diffusion equation is attained for $t\rightarrow\infty$ \cite{sgw21,gw22}, it equals the Bose--Einstein/Fermi--Dirac equilibrium distribution $n_\mathrm{eq}^\pm(\epsilon)$ 
\begin{equation}
n_\infty^\pm(\epsilon)=n_\mathrm{eq}^\pm(\epsilon)=\frac{1}{e^{(\epsilon-\mu)/T}\mp 1}\,,
 \label{Bose--Einstein}
\end{equation}
provided the ratio $v/D$ has no energy dependence, requiring the limiting value
%for the limit of time to infinity
$\lim_{t\rightarrow \infty}[-v\,(\epsilon,t)/D\,(\epsilon,t)] \equiv 1/T$. 
The chemical potential is $\mu\leq0$ in a finite Bose system, and $\mu=\epsilon_\mathrm{f}>0$ for a Fermi system. It appears as an integration constant in the solution of 
Eq.\,(\ref{boseq}) for $t\rightarrow\infty$. 
For an energy-dependent diffusion coefficient, the term $n\,\partial D/\partial \epsilon$  in Eq.\,(\ref{boseq}) 
ensures to attain the correct equilibrium limit for $t\rightarrow \infty$ \cite{gw22}.
%This result implies also that the equation with constant transport coefficients \eqref{eq:const_NBDE} has a Bose--Einstein distribution as
%its stationary solution. 

%In the limit of energy-independent transport coefficients, the nonlinear boson diffusion equation  
%for the occupation-number distribution $n(\epsilon,t)$
% per unit volume  
%becomes
%has the simple form
%\begin{equation}
%\frac{\partial n}{\partial t}=-v\,\frac{\partial}{\partial\epsilon}\Bigl[n\,(1+n)\Bigr]+D\,\frac{\partial^2n}{\partial\epsilon^2}\,.
% \label{bose}
%\end{equation}
%Again, the thermal equilibrium distribution $n_\mathrm{eq}$ is a stationary solution
%with $\mu\leq0$ and $T=-D/v$. 

In the simplified case of constant transport coefficients, the nonlinear diffusion equation
preserves during the time evolution the essential features of Bose--Einstein and Fermi-Dirac
statistics that are contained in the bosonic or fermionic quantum Boltzmann equation. 
{For constant transport coefficients, Eq.\,(1)} is one of the few nonlinear partial differential equations with a clear physical meaning that can be solved exactly through a nonlinear transformation \cite{gw18,gw22}. The resulting solution becomes
%For an initial distribution \( n_{{i}}(\epsilon) \), the solution for the time-dependent occupation-number distribution becomes
\begin{eqnarray}
%    n(\epsilon,t) =\pm T \frac{\partial}{\partial\epsilon}\ln{\mathcal{Z}(\epsilon,t)} \mp \frac{1}{2}= \pm T\frac{1}{\mathcal{Z}} \frac{\partial\mathcal{Z}}%{\partial\epsilon} \mp \frac{1}{2}\,,
        n(\epsilon,t) =\pm T\, \partial_\epsilon\ln{\mathcal{Z}(\epsilon,t)} \mp \frac{1}{2}= \pm \frac{T}{\mathcal{Z}}\, \partial_\epsilon \mathcal{Z} \mp
         \frac{1}{2}\,,
    \label{net} 
    \end{eqnarray}
 where the time-dependent partition function ${\mathcal{Z}(\epsilon,t)}$ obeys a linear diffusion equation which has a Gaussian Green's function $G(\epsilon,x,t)$.
  %  \begin{eqnarray}
 %   \frac{\partial}{\partial t}{\mathcal{Z}}(\epsilon,t) = D \frac{\partial^2}{\partial\epsilon^2}{\mathcal{Z}}(\epsilon,t)\,.
   % \label{eq:diffusionequation}
%\end{eqnarray}
The partition function can be written as an integral over the Green's function and an exponential function $F(x)$ 
%     \begin{eqnarray}
  %  \mathcal{Z}(\epsilon,t)= \int_{-\infty}^{+\infty} G(\epsilon,x,t)\,F(x)\,{d}x\,.
 %   \label{eq:partitionfunctionZ}
 %   \end{eqnarray}
%    $F(x)$ depends on 
which depends on
 the initial occupation-number distribution $n_\mathrm{i}$
according to
 \begin{eqnarray}
    F(x) = \exp\Bigl[ -\frac{1}{2D}\bigl( v x\pm 2v \int_0^x n_{{i}}(y)\,{d}y \bigr) \Bigr]\,.
       \label{ini}
\end{eqnarray}
%The definite integral over the initial conditions taken at the lower limit in Eq.\,(\ref{ini}) drops out in the calculation of $n(\epsilon,t)$ and can be replaced \cite{rgw20} by the indefinite integral
%with the integration constant set to zero without affecting the accuracy of the calculation. 
 %, such that
%\begin{eqnarray}
%    F(x) = \exp\Bigl[-\frac{1}{2D}\left( v x+2v A_{{i}}(x) \right)\Bigr]\,.
%    \label{eq:G(x)}
%\end{eqnarray}
%This replacement still provides the exact solution. 
%These modifications allow us to compute the partition function and the overall solution for the occupation-number distribution function Eq.\,(\ref{net}) analytically.
%, even in the presence of a singularity in the initial conditions, and with boundary conditions at the singularity $\epsilon = \mu < 0$ \cite{gw20}.

With the free Green's function $G\equiv G_\mathrm{free}$ of the linear diffusion equation, the physically correct solution is obtained for fermions \cite{bgw19}, but not for bosons due to the singularity in the infrared (IR): One has to consider the boundary conditions at the singularity $\epsilon = \mu \leq 0$ \cite{gw20}. They can be expressed as
 \(\lim_{\epsilon \downarrow \mu} n(\epsilon,t) = \infty\) \,$\forall$ \(t\). One obtains a vanishing partition function at the boundary \( \mathcal{Z} (\mu,t) = 0\), and the energy range is restricted to  $\epsilon \ge \mu$. This requires a new Green's function \cite{gw20,rgw20} 
%\( {F} (\epsilon,x,t) \) \cite{EqWorld} 
that equals zero at \(\epsilon = \mu\) $\forall \,t$. It can be written as
\begin{eqnarray}
    {G}_\mathrm{b} (\epsilon,x,t) = G_\mathrm{free}(\epsilon - \mu,x,t) - G_\mathrm{free}(\epsilon - \mu,-x,t),
    \label{eq:newGreens}
\end{eqnarray}
    and the partition function with this boundary condition becomes
     \begin{eqnarray}
    \mathcal{Z}_\mathrm{b} (\epsilon,t)= \int_{\mu}^{+\infty} G_\mathrm{b} (\epsilon,x,t)\,F(x)\,{d}x\,,
    \label{eq:partitionfunctionZ}
    \end{eqnarray}
which is equivalent to 
%$\mathcal{Z}_\mathrm{b} (\epsilon,t)= 
$\int_{0}^{+\infty} G_\mathrm{b} (\epsilon,x,t)\,F(x+\mu)\,{d}x$:
The function $F$ remains unaltered with respect to Eq.\,(\ref{ini}), but its argument is shifted by the chemical potential.
With the free Green's function for fermions, and the bounded Green's function for bosons, we calculate the time-dependent partition
function, which can be done analytically for sufficiently simple initial conditions, and obtain the occupation-number distributions via the basic nonlinear transformation Eq.\,(\ref{net}).

\section{Thermalization of quarks}
 \begin{figure}
 \begin{center}
 \includegraphics[scale=0.6]{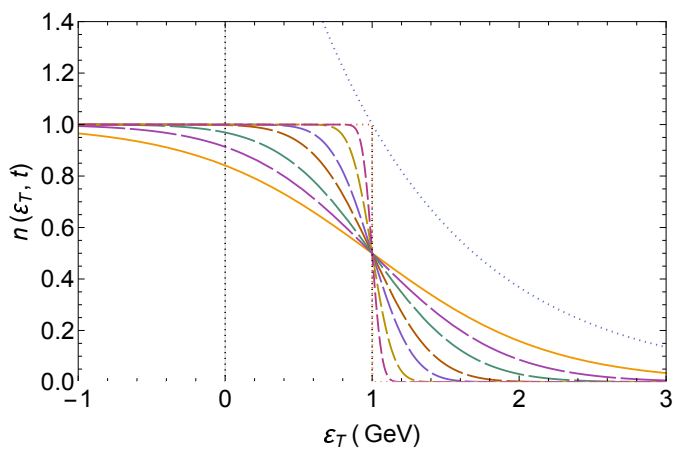}%
\caption{\label{fig1} {{Thermalization of quarks.} Analytical solutions of the fermionic nonlinear diffusion equation for quarks
show the time-dependent approach to the Fermi--Dirac equilibrium distribution (solid curve) with $T=-D/v=600$\,MeV at times $t=1\times10^{-3}, 5\times10^{-3},  0.02, 0.05, 0.15~\mathrm{and}~0.4$ fm/$c$ (increasing dash lengths).
The transport coefficients are $D=1.44$ GeV$^2$/fm, $v=-2.4$ GeV/fm, the fermionic equilibration time is $\tau_\mathrm{eq}^\mathrm{F}=4D/v^2=1$ fm/$c$. The dotted curve is the Maxwell--Boltzmann distribution.}}
\end{center}
\end{figure}
The model is first applied to the thermalization of valence quarks in central relativistic heavy-ion collisions such as Pb--Pb at $\sqrt{s_\mathrm{NN}}=5.02$\,TeV. For an initial condition taken as a $\theta$-function up to the Fermi energy, $n_\mathrm{i}(\epsilon)=\theta({\epsilon_\mathrm{f}}-\epsilon)$, the function $F(x)$, the partition function $\mathcal{Z}$ and the single-particle distribution function can be evaluated analytically \cite{bgw19}. Typical results
for the transport coefficients $D, v$ of Table\,1 are shown in Fig.\,\ref{fig1} for six timesteps, a fermionic equilibration time $\tau^\mathrm{F}_\mathrm{eq}=4D/v^2=1$\,fm/$c$ as derived in \cite{gw18}, and an equilibrium temperature of $T=-D/v=600$\,MeV. 

The time-dependent solutions at times beyond 0.1 fm/$c$ and the Fermi-Dirac equilibrium distribution at such a high initial temperature extend to negative energies, which is interpreted as antiparticle creation from the Fermi sea during the initial stages of the collision \cite{bgw19}. The initial temperature will then fall during expansion of the hot fireball until the hadronization temperature of about $T_\mathrm{H}\simeq160$\,MeV is reached, and light hadrons such as pions and kaons are formed.
\begin{table*}
\begin{center}
\hspace{4cm}
\caption{{Transport coefficients $D, v$ and equilibration times  $\tau_\mathrm{eq}$ for thermalization of quarks and gluons in central relativistic heavy-ion collisions, and of ultracold $^{39}$K atoms with Bose--Einstein condensate formation (scattering length $a=140\,a_0$).}}
%($T_\mathrm{i}=130$ nK, $T_\mathrm{f}=-D/v=32.5$ nK)\\
\vspace{.2cm}
\label{tab1}       % Give a unique label
\begin{footnotesize}
\begin{tabular}{rrrrrr}
\hline\noalign{\smallskip}
%\hline\noalign{\smallskip}
%&System&Energy&$D$\,(nK$^2$/ms)&$v$\,(nK/ms) &$\tau_\mathrm{eq}$\\
&System~~~~~~&Energy~~~&$D$~~~~~~~~&$v$~~~~~~~~~~&$\tau_\mathrm{eq}$~~~~~\\
%&&&$$~~~    (nK$^2$/ms)&$$~~~   (nK/ms) &\\
\noalign{\smallskip}\hline\noalign{\smallskip}
Quarks~~&$^{208}$Pb--$^{208}$Pb~~~~&5.02 TeV&1.44\,GeV$^2$/fm&$-2.4$\,GeV/fm~~&1.0\,fm/$c$\\
Gluons~~&$^{208}$Pb--$^{208}$Pb~~~~&5.02 TeV&1.20\,GeV$^2$/fm&$-2.0$\,GeV/fm~~&~~0.13\,fm/$c$\\
Atoms~~&$^{39}$K~~~~~~~~~~&195 nK~~&~~0.08\,nK$^2$/ms~~&$-0.00246$\,nK/ms&600 ms~~~\\ 
\noalign{\smallskip}\hline
\end{tabular}
\end{footnotesize}
\end{center}
\end{table*}
 \begin{figure}
 \begin{center}
 \includegraphics[scale=0.57]{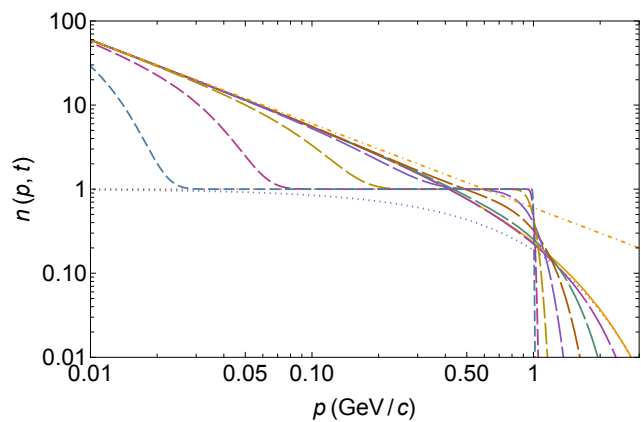}%
\caption{\label{fig2} {{Thermalization of gluons.} Analytical solutions of the nonlinear boson diffusion equation (NBDE) for gluons
with boundary conditions at the singularity show the time-dependent approach to the Bose--Einstein equilibrium distribution (solid curve) with $T=600$ MeV at times $t=2\times10^{-5},2\times10^{-4}, 2\times10^{-3}, 0.012, 0.04, 0.12~\mathrm{and}~0.4$ fm/$c$ (increasing dash lengths).
The transport coefficients are $D=1.2$ GeV$^2$/fm, $v=-2$ GeV/fm, the bosonic equilibration time is $\tau_\mathrm{eq}^\mathrm{B}=4D/(9v^2)=0.13$ fm/$c$. The dotted curve is the Maxwell--Boltzmann, the dot-dashed line the Rayleigh--Jeans distribution.}}
\end{center}
\end{figure}
\section{Thermalization of massless gluons}

To account for the local thermalization of gluons, the boundary conditions at the singularity must be considered. For massless gluons with
%$\epsilon=bf{p}=p$ and $\theta$-function initial conditions 
$\epsilon=p$ and $\theta$-function initial conditions 
\cite{mue00} $n_\mathrm{i}=n_\mathrm{i}^0\,\theta(1-p/Q_\mathrm{s})$ with the gluon saturation momentum $Q_\mathrm{s}\simeq1$\,GeV and an average initial occupation $n_\mathrm{i}^0$, the function $F(x)$ can be evaluated in case of inelastic gluon scatterings with vanishing chemical potential $\mu=0$ as \cite{gw22}
\begin{eqnarray}
	\label{fxmu1}
		&F\,(x)=\exp[-v\,x/(2D)\\ \nonumber 
		-&(v\,n_\mathrm{i}^0/D)(Q_\mathrm{s}-x)\,\theta\,(x-Q_\mathrm{s})+x]\,.
	%F\,(x)=\exp\left[-\varv\,(x+\mu)/(2D)\right]\qquad\qquad\\ \nonumber
	%\times\exp\left[-(\varv\,n_\mathrm{i}^0/D)(Q_\mathrm{s}-x-\mu)\,\theta\,(x+\mu-Q_\mathrm{s})+x\right]\,.
\end{eqnarray}
With the bounded Green's function Eq.\,(\ref{eq:newGreens}), the partition function and the gluon distribution function can be evaluated, 
%{by solving the nonlinear boson diffusion equation (NBDE)}
as shown in Fig.\,\ref{fig2} for $n_\mathrm{i}^0=1$ corresponding to an overoccupied situation \cite{jpb12} with the diffusion and drift coefficients of Table\,1. This results in a bosonic equilibration time $\tau^\mathrm{B}_\mathrm{eq}=4D/(9v^2)=0.13$\,fm/$c$ as derived in \cite{gw18} at $p=Q_\mathrm{s}$, and an equilibrium temperature of $T=-D/v=600$\,MeV. 

The thermalization is faster as in case of quarks, because it is not hindered by Pauli's principle. The difference is enhanced by the colour factors $C_\mathrm{A}=3$ for gluons, and $C_\mathrm{f}=4/3$ for quarks. 

{For gluons, the assumption of constant transport coefficients seems critical when viewed from the aspect of small-angle scattering equations \cite{jpb12,jpbl15,blmt17}. 
However, when inserting physical timescales in the numerical calculations presented for inelastic gluon scattering in Ref.\,\cite{blmt17}, I obtain very similar analytical results for $D=\mathrm{const}$. There is presently no evidence that momentum-dependent diffusion coefficients are mandatory in gluon thermalization. }
 More detailed results including thermalization in an underoccupied system with the initial average occupation below the critical value \cite{jpb12} $r_\mathrm{c}=0.154$ for $\theta$-function initial conditions and inelastic gluon scatterings with $\mu=0$ are presented in Ref.\,\cite{gw22}.

\section{Thermalization of ultracold atoms}
 \begin{figure}
 \begin{center}
 \includegraphics[scale=0.58]{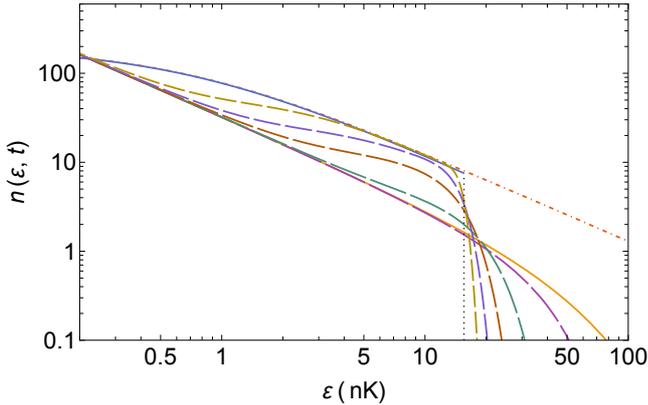}%
\caption{\label{fig3} {{Thermalization of ultracold potassium atoms.}  The analytical solutions of the nonlinear boson diffusion equation (NBDE) for ultracold  $^{39}$K atoms are displayed as functions of time.
% for $\mu=0$. 
The initial state is a Bose--Einstein distribution with 
%$\mu_\mathrm{i} = -0.67$ nK and 
a cut at $\epsilon_\mathrm{i}=15.56$\,nK, upper solid curve. The initial temperature is $T_\mathrm{i} = 130$\,nK, the final temperature $T_\mathrm{f} = -D/v = 32.5$\,nK (lower solid curve). Exact solutions of the NBDE are obtained as a finite series expansion with $k_\mathrm{max}=T_\mathrm{i}/T_\mathrm{f}=4$.
	The time evolution of the single-particle occupation-number distributions is shown at $t =8, 30, 100, 400$, and $4000$\, ms (increasing dash lengths). The transport coefficients are $D = 0.08$\,(nK)$^2$/ms,\,$v = -0.00246$\,nK/ms, the condensate formation time is $\tau_\mathrm{eq}^\mathrm{c}=f\,D/v^2=600$ ms, with a proportionality factor $f\simeq0.045$. The dot-dashed line is the Rayleigh--Jeans power law in the initial distribution.}}
	\end{center}
\end{figure}
\begin{figure}
\begin{center}
 \includegraphics[scale=0.57]{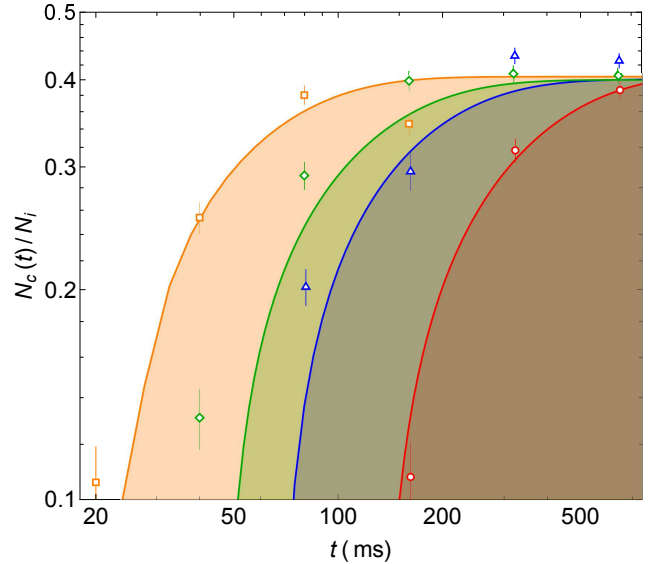}%
\caption{\label{fig4} {{Time-dependent condensate fractions for an equilibrating 3D potassium Bose gas.}  The calculated condensate fractions based on the analytical solutions of the NBDE are displayed as functions of time (solid curves) and compared to Cambridge data \cite{gli21} (symbols, with error bars reflecting experimental fitting errors.). In the experiment, ultracold $^{39}$K atoms in a box trap are subjected to a rapid quench from $T_\mathrm{i} = 130$ nK to $T_\mathrm{f} = 32.5$ nK. Model results for an initial chemical potential
$\mu_\mathrm{i} = -0.67$ nK and a quench at $\epsilon_\mathrm{i} = 15.56$ nK are shown in a double-log plot for scattering lengths $a/a_0=$140 (circles), 280 (triangles), 400 (diamonds) and 800 (squares). The transport coefficients given in Table\,1 for  
$a/a_0=140$ are scaled with the scattering lengths.}}
\end{center}
\end{figure}

Turning to the thermalization of ultracold atoms via number-conserving elastic collisions, the initial nonequilibrium distribution that is encountered in evaporative cooling --- or following a sudden energy quench \cite{gli21} --- can be represented as a Bose--Einstein equilibrium distribution with an energy cutoff at $\epsilon=\epsilon_\mathrm{i}$. The solution of the nonlinear boson diffusion equation (NBDE) for the combined initial- and boundary value problem \cite{gw20,rgw20} can still be calculated analytically, with the time-dependent partition function expressed as an infinite series
\begin{eqnarray}
{\mathcal{Z}}(\epsilon,t) &= \sqrt{4 D t} \, \exp\Bigl(-\frac{\mu}{2 T_\mathrm{f}}\Bigr)\qquad\qquad\\ \nonumber
 &\times\sum_{k=0}^{\infty} \dbinom{\frac{T_\mathrm{i}}{T_\mathrm{f}}}{k} (-1)^k \times
f_k^{{T}_\mathrm{i},{T}_\mathrm{f}} \,(\epsilon,t)\,.
\label{partfct}
\end{eqnarray}
The analytical expressions for $f_k^{{T}_\mathrm{i},{T}_\mathrm{f}} \,(\epsilon,t)$ are combinations of exponentials and error functions, they are given explicitly in
 Ref.\,\cite{rgw20}, and similarly, for the derivative of ${\mathcal{Z}}(\epsilon,t)$. The series terminates at $k_\mathrm{max}$ in case $T_\mathrm{i}/T_\mathrm{f}$ is an integer, which I use here to compute the exact solution for the thermalization and condensate formation of ultracold $^{39}$K atoms that has recently been investigated experimentally in Ref.\,\cite{gli21}, where $k_\mathrm{max}=T_\mathrm{i}/T_\mathrm{f}=4$. 
 
The resulting time-dependent distribution functions following a quench at $\epsilon_\mathrm{i}=15.56$ nK of an initial thermal distribution of ultracold $^{39}$K atoms with $T_\mathrm{i}=130$\,nK corresponding to a removal of 77\% of the particles, 97.5\% of the total energy \cite{gli21}, and a final temperature $T_\mathrm{f}=32.5$\,nK that is below the critical temperature are shown in Fig.\,\ref{fig3} at five timesteps between 8\,ms and 4000\,ms, with transport coefficients $D, v$ from Table\,1 and an equilibration time of 600\,ms, which equals the condensate formation time: The condensate fraction reaches its equilibrium value within this time. The calculations are done for an interaction strength that corresponds to a scattering length $a=140\,a_0$, with $a_0=0.529\times10^{-10}$\,m the Bohr radius. 
%This is close to the physical scattering length of $^{39}$K atoms, which is $138.49\,a_0$. 
Details are presented in Ref.\,\cite{kgw22}.

With the analytical solutions of the nonlinear boson diffusion equation, we can now calculate the time-dependent condensate fraction from particle-number conservation as $N_\mathrm{c}(t)/N_\mathrm{i}=1-N_\mathrm{th}(t)/N_\mathrm{i}$, where $N_\mathrm{i}$ is the initial particle number just after the quench, and $N_\mathrm{th}(t)$ the integral of the time-dependent solution of the NBDE once the chemical potential has reached the value $\mu=0$ such that condensate formation starts, 
\begin{equation}
N_\mathrm{th}(t)=\int_0^\infty g(\epsilon)\,n(\epsilon,t)\,{d}\epsilon\,.
\label{Nth}
\end{equation}
For a threedimensional box trap as in the Cambridge potassium experiment \cite{gli21}, the density of states is $g(\epsilon)=g_0\sqrt{\epsilon}$, and with the analytical nonequilibrium solutions $n(\epsilon,t)$ of Eqs.\,(\ref{net}),(\ref{partfct}), the integral is evaluated numerically 
%using tt{C++} 
since presently no analytical solution is known, except for $t\rightarrow\infty$.

The Cambridge experiment has been performed for various scattering lengths, and here the analytical NBDE-results are compared in Fig.\,\ref{fig4} with data for $a/a_0=140, 280, 400$ and 800, where the transport coefficients start with the values given in Table\,1 for $a=140 a_0$, and are scaled with the scattering lengths rather than the cross sections due to the emerging coherence between the highly occupied IR-modes \cite{sto97,dwg17}.

{As in the high-energy case, the transport coefficients are assumed to be constant in the present cold-atom model calculations. An energy- or time-dependence of the diffusion coefficient would be required if the measured start of condensate formation was gradual, and not abrupt as is the case for constant diffusion coefficient. 

There are indeed signs of a gradual onset in the historical $^{23}$Na MIT data \cite{miesner_bosonic_1998}, but not in the more recent $^{87}$Rb data, which have detailed error bars \cite{kdg02}. The current $^{39}$K deep-quench Cambridge data  \cite{gli21} show some indication of a gradual onset only at small scattering length $a=140a_0$  \cite{kgw22}.
%, but overall, results with constant diffusion coefficients yield reasonable agreement with the data. 
Hence, the approximation $D=\mathrm{const}$ appears permissible also for cold atoms.}

For $a=140\,a_0$, data and model result almost coincide {beyond the onset}, whereas for the other scattering lengths, some deviations {larger than} the error bars are observed. In particular, the $1/a$-dependence of the timescales is not well fulfilled in the data for $a=400\,a_0$. But the overall agreement is rather satisfatory, 
the equilibrium limit $N^\mathrm{eq}_\mathrm{c}/N_\mathrm{i}=1-g_0T_\mathrm{f}^{3/2}\zeta(3/2)\Gamma(3/2)/N_\mathrm{i}$ which equals the experimental value of $0.40(5)$ \cite{gli21} is approached within the condensate formation times. 
The infrared power index $\alpha_\mathrm{IR}=1.08(9)$ that the experimentalists have extracted from their data is consistent with the NBDE calculation.
However, in the present formulation of the nonlinear diffusion model, thermalization proceeds continuously in time without nonthermal fixed points \cite{dwg17}.

These results confirm the validity and usefulness of the analytical nonlinear diffusion model and encourage further applications such as to the thermalization of fermionic cold atoms, but also in different areas of physics like reheating in inflationary cosmology.

\acknowledgments

Contributions of current or former Heidelberg students Thomas Bartsch, Johannes H\"olck, Anton Kabelac, Niklas Rasch and Alessandro Simon
to applications of the nonlinear diffusion model are gratefully acknowledged, as well as discussions with Thomas Gasenzer, Zoran Hadzibabic, Pascal Naidon, {and questions of the referee}.
I thank Emiko Hiyama for her hospitality at Tohoku University, Sendai, and RIKEN, Wako, where this work was completed with the support of {JSPS-BRIDGE fellowship BR200102}.

%\begin{thebibliography}{0}
%\bibliographystyle{eplbib}

%\bibliography{gw_22.bib}% common bib file
%!TEX encoding = UTF-8 Unicode\input gw_epl_22.bbl
%\end{thebibliography}


\end{document}